\documentclass[12pt]{article}

\usepackage{graphicx,color}
\usepackage{amsmath}
\usepackage{amsfonts}
\usepackage{caption}
\usepackage{amsthm}
\newtheorem{definition}{Definition}
\newtheorem{lemma}{Lemma}
\newtheorem{corollary}{Corollary}
\newtheorem{theorem}{Theorem}

\usepackage{times}
\usepackage{bm}
\usepackage{natbib}

\usepackage[plain,noend]{algorithm2e}

\makeatletter
\renewcommand{\algocf@captiontext}[2]{#1\algocf@typo. \AlCapFnt{}#2} % text of caption
\def\@algocf@capt@plain{top}
\renewcommand{\algocf@makecaption}[2]{%
  \addtolength{\hsize}{\algomargin}%
  \sbox\@tempboxa{\algocf@captiontext{#1}{#2}}%
  \ifdim\wd\@tempboxa >\hsize%     % if caption is longer than a line
    \hskip .5\algomargin%
    \parbox[t]{\hsize}{\algocf@captiontext{#1}{#2}}% then caption is not centered
  \else%
    \global\@minipagefalse%
    \hbox to\hsize{\box\@tempboxa}% else caption is centered
  \fi%
  \addtolength{\hsize}{-\algomargin}%
}
\makeatother

\begin{document}

\begin{center}
	\section*{Optimal Design Emulators: A Point Process Approach}
	\vspace{-.25cm}
	
	\textbf{Matthew T. Pratola}$^{1,3}$,
	\textbf{C. Devon Lin}$^{2,4}$,
	\textbf{Peter F. Craigmile}$^{1,5}$

	$^1$
	Department of Statistics,
	The Ohio State University,
	Columbus, OH 43210, USA
	
	$^2$
	Queen's University,
	Department of Mathematics and Statistics,
	Jeffery Hall, University Avenue,
	Kingston, ON K7L 3N6, Canada
	
	$^3$\verb_mpratola@stat.osu.edu_ \;,
	$^4$\verb_devon.lin@queensu.ca_ \;,
	$^5$\verb_pfc@stat.osu.edu_ \;

	\textit{Last updated \today}
	\vspace{-.2cm}
\end{center}

\begin{abstract}
\noindent
Design of experiments is a fundamental topic in applied statistics with a long history. Yet its application is often limited by the complexity and costliness of constructing experimental designs, which involve searching a high-dimensional input space and evaluating computationally expensive criterion functions.  
In this work, we introduce a novel approach to the challenging design problem. We will take a probabilistic view of the problem by representing the optimal design as being one element (or a subset of elements) of a probability space. Given a suitable distribution on this space, a generative point process can be specified from which stochastic design realizations can be drawn. In particular, we describe a scenario where the classical entropy-optimal design for Gaussian Process regression coincides with the mode of a particular point process.  We conclude with outlining an algorithm for drawing such design realizations, its extension to sequential designs, and applying the techniques developed to constructing  designs for Stochastic Gradient Descent and Gaussian process regression. 
\end{abstract}

%\begin{keywords}
Big data; Design of experiments; Determinantal point process; Exchange algorithm; Gaussian process; Stochastic gradient descent; Sequential design.
%\end{keywords}

% ======================================================================
\newpage
\section{Introduction}
\label{sec:introduction}
% Introduction

Optimal design of experiments (DOE) is a fundamental topic in statistics %with a long history
\citep[e.g.][]{Fedorov:1972,Silvey:1980,Kiefer:1985,Atkinson:2007,Dean:etal:2017}.  More recently, DOE has seemed to attract less attention from theoreticians, methodologists, and practitioners.  One reason may be due to an increased reliance on observational data, but this ignores current challenges in  statistical analysis.  Datasets are becoming larger and more complex.  Statistical models are more complicated.  Uncertainty quantification is more challenging.  As the challenges increase, DOE should play an important role in modern statistical analyses.

Typically DOE targets a particular aspect of a model which is deemed ``important'', such as the prediction error in a spatial model, or the variance of parameters in a regression model.  Given $n$ input settings ${\boldsymbol\xi}_1,\ldots,\boldsymbol{\xi}_n$ drawn from some compact set $\chi\subset\mathbb{R}^d$, a design criterion $\mathcal{L}(\boldsymbol{\xi}_1,\ldots,\boldsymbol{\xi}_n)$ is specified, and the $n$-run optimal design involves finding the input settings $\boldsymbol{\Xi}_n$ that minimize this criterion:
\begin{align}
\label{eqn:designproblem}
\boldsymbol{\Xi}_n=\arg\min_{\boldsymbol{\xi}_1,\ldots,\boldsymbol{\xi}_n}\mathcal{L}(\boldsymbol{\xi}_1,\ldots,\boldsymbol{\xi}_n).
\end{align}
Assuming $\chi$ is a discretized candidate set of cardinality $N$, the number of possible designs to explore is ${N \choose n}.$ The {\em Exchange Algorithm} \citep{Fedorov:1972} is the most popular approach to solving this problem, which performs one-at-a-time updates to the design.  Unfortunately, solving (\ref{eqn:designproblem}) is notoriously difficult due to the large number of possible designs and the multi-modality of the optimization problem. More recently, modern  algorithms such as particle-swarm methods and simulated annealing \citep{morris:mitchell:1995,jin:etal:2008,Chen:etal:2013} have been applied, but these can be difficult to implement reliably in modern settings \citep[e.g.][]{nguyen:etal:2019}.

When the dimensionality of the input space is high or the number of design points is large, solving (\ref{eqn:designproblem}) remains practically infeasible.  This is because constructing a designed experiment involves searching the $d$-dimensional input space $\chi\subset\mathbb{R}^d$, and, for each plausible solution, calculating an optimality criterion which can itself be computationally expensive.  This challenging problem has typically been made tractable by changing the optimality criterion to one based on a simplified model that is more computationally amenable. 

The goal of this work is to introduce a novel approach to constructing optimal designs
for Gaussian process (GP) regression models, which have broad applications in spatial statistics, computer experiments and statistical/machine learning \citep{sack:welc:mitc:wynn:1989,Furrer:2006:Covariance,Cressie:2008:Fixed,Banerjee:2008:Gaussian,Higdon:etal:2008,Guhaniyogi:2011:Adaptive,Sang:2011:Covariance,Katzfuss:2013:Bayesian,Pratola:etal:2014,Pratola:etal:2014b,
Gramacy:Apley:2015,katzfuss2014parallel}. We take a probabilistic view of the problem that has recently become more popular.  %is more general than the traditional probabilistic view previously discussed in the DOE literature \citep{Kiefer:1985,Muller:2007}.  
For instance, \cite{franco2008exploratory} and \cite{franco:2008} made use of Strauss processes motivated by the popular space-filling heuristic, and \cite{mazoyer:2019} take a probabilistic perspective on space-filling designs, focusing particularly on low-dimensional projection properties.  In our work, we outline a probabilistic approach having an explicit connection to statistical modeling and DOE theory.  The idea is to represent the collection of points that form an optimal design as being one element (or a subset of isometrically equivalent elements) of a stochastic process defined on the space of point patterns.  By specifying an appropriate {\bf generative point process} (PP) for this distribution, we introduce the idea of an {\em optimal design emulator}, where the classical optimal design solution typically coincides with the mode of this generative stochastic process.  Since the generative process can be specified in terms of a low-dimensional parameter space, constructing an optimal design reduces to drawing a realization of this process given appropriately tuned parameter settings of the process rather than performing a difficult optimization problem.  This approach lends itself to a wide selection of Markov Chain Monte Carlo (MCMC) tools for efficient computation, and also gives a measure of how optimal the design drawn actually is.
As such, our work draws a connection between PP models and optimal designs -- particularly for GP's -- while taking a distinctively Bayesian perspective on the design problem.

\subsection{GP Regression and Design}
\label{sec:gpintro}
The GP regression model is used extensively in modern applications as a model of
an unknown process $f({\bf x})$ observed at continuous inputs ${\bf x} \in \mathbb{R}^d$.  The continuous inputs ${\bf x}$ represent the input settings where our process may be observed and/or predicted.  In contrast, the discrete, countable, set of {\em design candidates} is represented as $\chi\subset\mathbb{R}^d,$ from which $n$-run experimental designs $\boldsymbol{\Xi}=\lbrace\boldsymbol{\xi}_1,\ldots,\boldsymbol{\xi}_n\rbrace$ may be constructed, where $\boldsymbol{\xi}_i\in\chi,\ i=1,\ldots,n.$   The process $f({\bf x})$ may or may not be observed with noise $\epsilon({\bf x})$, leading to a model for the observations $y({\bf x}),$ given by
\[y({\bf x})=f({\bf x})+\epsilon({\bf x}).\]
The error term $\epsilon({\bf x})$ is often assumed to be independent and identically distributed (i.i.d.) normal with mean zero and error variance $\sigma_\epsilon^2$ (sometimes called the nugget). % Examples include spatial and spatial-temporal modeling \citep{Cres:93,Furrer:2006:Covariance,Paciorek:2007:Bayesian,Cressie:2008:Fixed,Banerjee:2008:Gaussian,Kaufman:2008:Covariance,Guhaniyogi:2011:Adaptive,Sang:2011:Covariance,Lindgren:2011:An-explicit,Katzfuss:2013:Bayesian,katzfuss2014parallel} as well as emulating complex response surfaces in computer experiments \citep{sack:welc:mitc:wynn:1989,kenn:ohag:2001,Vernon:etal:2010,Higdon:etal:2008,Gramacy:Lee:2008,Gramacy:Apley:2015,Pratola:etal:2014,Higdon:etal:2013,Higdon:etal:2004,Joseph:Melkote:2009,Oakl:O'Ha:2002,Goldstein:Rougier:2009,Goldstein:Rougier:2006}
In the simplest case, the unknown process $f({\bf x})$ is modeled as a stationary GP with mean $E[f({\bf x})]=\mu({\bf x})$ and covariance $\text{Cov}(f({\bf x}),f({\bf x}^\prime))=\sigma^2 c({\bf x},{\bf x}^\prime)$ at input settings ${\bf x},{\bf x}^\prime\in\mathbb{R}^d$
%
%\[f({\bf x}) \sim N\left(\mu({\bf x}),\sigma^2c(\cdot,\cdot)\right)\]
%
where the mean model is $\mu({\bf x})$, the process scale is $\sigma^2$, and $c({\bf x},{\bf x}^\prime)$ is a positive
definite correlation function.  The choice
of mean function can be as simple as a constant or can include
covariates that are related to $f$.  The popular choice of an isotropic Gaussian
correlation function \citep[e.g.][]{sack:welc:mitc:wynn:1989}, 
\begin{align}
\label{eqn:corrfn}
c({\bf x},{\bf x}^\prime )=\rho^{\vert\vert {\bf x}-{\bf x}^\prime\vert\vert^2},
\end{align} 
assumes a smooth,
continuous, and infinitely differentiable response, where $\rho$ is the correlation parameter of the GP.  Given $n$ observations ${\bf Y}=\left(y({\bf x}_1),\ldots,y({\bf x}_n)\right)$, and assuming $\mu({\bf x})=0$ for all ${\bf x},$ the GP model is % likelihood $L(\sigma^2,\rho | {\bf Y})$ is
\begin{align}
\label{eqn:gpmodel}
{\bf Y} | \sigma^2,\rho \sim N_n({\bf 0},\sigma^2{\bf R}+\sigma_\epsilon^2{\bf I}),
\end{align}
where ${\bf R}_{ij}=c({\bf x}_i,{\bf x}_j)$ is the $(i,j)$ entry of the correlation matrix ${\bf R}$.

In the setting of spatial statistics or computer experiments, the two most popular model-based design criteria are the {\em integrated mean squared prediction error (IMSPE) optimal designs}, $\mathcal{L}=\int_{\bf x}\left(Y({\bf x})-E[Y({\bf x})\vert\boldsymbol{\xi}_1,\ldots,\boldsymbol{\xi}_n]\right)^2d{\bf x}$ and the {\em entropy optimal designs}, $\mathcal{L}=E\left[\hbox{log}(f_{\bf Y})\right]$ where $f_{\bf Y}$ is the usual multivariate Gaussian density corresponding to  (\ref{eqn:gpmodel}).  %IMSPE optimal designs are useful as they minimize the error in out-of-sample predictions.  Entropy optimal designs provide improved estimates of the GP correlation parameter, $\rho,$ which can be important for accurately quantifying prediction uncertainties, interpreting which variables are active in a variable selection problem \citep{Morris:etal:1993,Linkletter:etal:2006}, or improved estimation of the variogram \citep{Cres:93}.  
Both the IMSPE and entropy-based criteria involve $\mathcal{O}(n^3)$ operations on the potentially large $n \times n$ correlation matrix ${\bf R}$. The IMSPE criterion additionally involves integration over the design domain (although see \cite{spock:pilz:2010,gauthier:pronzato:2017} for possible workarounds). %, which makes an already challenging optimization problem even more difficult.  
%An alternative approach is to use model-robust designs that are geometrically motivated, such as the Latin Hypercube Sampling (LHS) designs \citep{Mckay:etal:1979}, or other ``space-filling'' designs such as maximin distance designs \citep{Nychka:etal:2015,r:lhs}.
An alternative approach is to use model-robust designs that are geometrically motivated.  For example, Latin Hypercube Sampling (LHS) designs \citep{Mckay:etal:1979} can be space-filling, however caution is needed to ensure we do not generate undesirable LHS designs that occur purely on the diagonal.  There are also other ``space-filling'' designs such as maximin distance designs \citep{Nychka:etal:2015,r:lhs}.
%LHS designs are popular model-free designs that are colloquially described as ``space-filling'' designs, and are theoretically justified as variance-reduction designs \citep{Mckay:etal:1979} while also having theoretical connections to Gaussian Process (GP) regression models \citep{Johnson:Moore:Ylvisaker:1990}.
Maximin distance designs were found to be the limiting form of entropy optimal designs for GP regression as the correlation $\rho$ decays to $0$ \citep{Johnson:Moore:Ylvisaker:1990}.  Similar to LHS and maximin designs, both IMSPE and entropy designs empirically lead to designs exhibiting space-fillingness. %, that is the chosen design points tend to spread out over the design region thereby filling-in any empty space, 
%such as the LHS sample shown in Figure \ref{fig:lhs:pp}(c). 
However, LHS and maximin designs are usually more amenable in terms of computational cost. In practice, a combined criterion is often used, such as the space-filling LHS implemented in the popular \texttt{R} package \texttt{fields} \citep{Nychka:etal:2015}.

\subsection{Point Processes}
\label{sec:ppintro}
The statistical study of {\em point patterns} attempts to classify what type of pattern a given, observed, set of points exhibits.  Such point patterns can be classified as random, clustered or regular.  Applications are numerous in the area of ecology where often the spatial location and spatial density of an object being studied is to be inferred and possibly also used for prediction. 
A visualization of this idea comparing random, clustered and LHS designs is shown in the Supplementary Material.
 
Let $\boldsymbol{\Xi}=\lbrace \boldsymbol{\xi}_1,\ldots,\boldsymbol{\xi}_n\rbrace$ represent a point pattern of cardinality $n$ and let $Z({\bf x})$ represent a stochastic point process defined on all ${\bf x}\in\mathbb{R}^d$.  Such a process assigns a probability measure $\mathcal{F}_Z: \boldsymbol{\Xi}\subseteq \mathcal{D}\rightarrow [0,1],$ where $\mathcal{D}\subset\mathbb{R}^d$ could be a continuous subset of $\mathbb{R}^d,$ or some discrete, countable subset of cardinality $\vert\mathcal{D}\vert=N.$  In the former interpretation, the PP can be viewed as assigning the probability that a point will be realized in some infinitesimal region $d{\bf x}$ about ${\bf x}.$  In the latter interpretation, one can view the PP realization as an $N$-vector such that $n$ entries take the value 1, indicating presence of some element $\boldsymbol{\xi}\in\mathcal{D}$ in the point pattern, and $N-n$ entries taking the value 0 indicating absence.  In either setting, 
the point pattern is described by the number, $n,$ of points making up the realization, and the location of these points, here denoted by the $\boldsymbol{\Xi}=\lbrace\boldsymbol{\xi}_1,\ldots\boldsymbol{\xi}_n\rbrace$ \citep{geyer1994simulation,Diggle:2005,Lavancier2015}.  In this work, we focus on the discrete case.

%%Typical applications of PP modeling are in spatial statistics \citep[e.g.][]{Diggle:2005} where the locations are coordinates in a subregion of $\mathbb{R}^2,$ but we consider more generally the location of points in a subregion of $\mathbb{R}^d.$  
We assume simple PP models, that is at any given location ${\bf x}$ we will only ever realize at most one point at ${\bf x}.$  We also assume stationary PP models throughout, although this is not necessary in general.
The simplest PP model is the Poisson \citep[e.g.][]{daley2003introduction}, where the probability of a point in the domain $\mathcal{D}$ belonging to the point pattern is given by the Poisson distribution with rate parameter given by the {\em first order intensity}, $\lambda_Z,$ defined as
%
%\vspace{-0.25cm}
\begin{eqnarray}
\label{eqn:intensity1}
	\lambda_Z({\bf x})
	&=&
%	\lim_{|d {\bf x}| \to 0} \frac{E(Z(d {\bf x}))}{|d {\bf x}|},
	E(Z({\bf x})),
\end{eqnarray}
which is simply the average number of points one expects to be generated at ${\bf x}$ under the Poisson assumption.  Furthermore, due to the memoryless property of the Poisson model, the existence of a point in some arbitrary Borel-measurable subregion $B$ of $\mathcal{D}$ is independent of the existence of a point in some arbitrary Borel-measureable subregion $B^\prime$ of $\mathcal{D}.$  For this reason, the Poisson model is referred to as a point process model of complete %spatial 
randomness.

Besides the complete random point patterns of the Poisson model, there are two additional possibilities: {\em clustered point patterns} and {\em regular point patterns}. These cases necessarily require additional parameterization to capture the form of non-independence of points that give rise to clustered or regular point patterns.  
The straightforward extension of the Poisson model is to introduce the so-called {\em second order intensity },
$\lambda_{2,Z}$, defined as
%
%\vspace{-0.25cm}
\begin{eqnarray}
\label{eqn:intensity2}
	\lambda_{2,Z}({\bf x}, {\bf x}')
	\;=\;
%	\lim_{|d {\bf x}| \to 0, |d {\bf x}'| \to 0} \frac{E(Z(d {\bf x})Z(d {\bf x}'))}{|d {\bf x}| |d {\bf x}'|}.
	E(Z({\bf x})Z({\bf x}')),
\end{eqnarray}
%
%The second-order intensity function
which simply captures the expected number of points that will jointly co-occur at ${\bf x}$ and ${\bf x}^\prime.$
The properties described by the intensities $\lambda_z, \lambda_{2,Z}$ provide one way of classifying the point process
%One can also classify the type of PP using %these intensity functions and the first order intensities 
%(\ref{eqn:intensity1}),(\ref{eqn:intensity2}) by essentially considering the covariance properties of the process
~\citep{Diggle:2005}:
%
%\vspace{-0.25cm}
\begin{eqnarray*}
	\lambda_{2,Z}({\bf x},{\bf x}^\prime) &<& \lambda_Z({\bf x})\lambda_Z({\bf x}^\prime) \Rightarrow \text{ regular point pattern},\\
	\lambda_{2,Z}({\bf x},{\bf x}^\prime) &=& \lambda_Z({\bf x})\lambda_Z({\bf x}^\prime) \Rightarrow \text{ complete spatial randomness},\\
	\lambda_{2,Z}({\bf x},{\bf x}^\prime) &>& \lambda_Z({\bf x})\lambda_Z({\bf x}^\prime) \Rightarrow \text{ clustered point pattern.}
\end{eqnarray*}
That is, $\lambda_{2,Z}({\bf x},{\bf x}^\prime)>\lambda_Z({\bf x})\lambda_Z({\bf x}^\prime)$ indicates that it is relatively more likely for points to co-occur than under the Poisson model, which generates clustered point patterns.  Similarly, $\lambda_{2,Z}({\bf x},{\bf x}^\prime)<\lambda_Z({\bf x})\lambda_Z({\bf x}^\prime)$ indicates that it is relatively less likely for points to co-occur than under the Poisson model, which generates regular point patterns.

%As alluded to in Section \ref{sec:pointpatterns}, the PP models that interest us are those that generate point patterns that fall in the regular class.  We will, in particular, consider PP models that are linked to optimal designs for our regression model of interest, the stationary Gaussian Process outlined in Section \ref{sec:introduction}.  The first hint at this connection comes from examining the LHS design, which is often used in GP regression, in further detail.

In Section 2 we will briefly review entropy-optimal designs for GPs and motivate that they belong to the class of regular point patterns.  In Section 3 we introduce a particular regular point process model, the {\em determinantal point processes} (DPPs) \citep{Lavancier:2015:Determinantal}, and show how these repulsive point process models are connected to the stationary GP model outlined in Section \ref{sec:gpintro}.  
We then introduce the concept of a \textit{design emulator} -- a PP model that assigns a probability measure to the space of possible design point patterns such that the classical optimal design coincides with the mode -- and
devise an efficient algorithm that can be used to sample entropy optimal designs for GPs from the emulator.  
Section 4 explores examples of our design emulator applied to the popular stochastic gradient descent algorithm and for sequential GP regression. We conclude in Section 5. All proofs of results are presented in the Appendix.

% ======================================================================

\section{Entropy Optimal Designs for Gaussian Processes}%Point Processes and Optimal Designs}
\label{sec:ppod}
Let us consider the simplest case of our GP regression model defined in Section \ref{sec:gpintro}, with a mean trend of $\mu({\bf x})=0$ and noise-free observations, i.e. $\sigma_\epsilon^2=0$.  The GP model in this case is commonly used in computer experiments and in spatial statistics where a ``nugget'' term is not required.  In this setting, the maximum entropy criterion for an $n$-run design, $\boldsymbol{\xi}_1,\ldots,\boldsymbol{\xi}_n,$ drawn from the discrete candidate set $\chi\subset\mathbb{R}^d$ can be shown to reduce to \citep{Shewry:Wynn:1987}
\begin{eqnarray}
\label{eqn:entopt}
\mathcal{L}(\boldsymbol{\xi}_1,\ldots,\boldsymbol{\xi}_n) &=& -\mbox{det}({\bf R}),
\end{eqnarray}
where the $n\times n$ correlation matrix ${\bf R}$ has entries ${\bf R}_{ij}=c(\boldsymbol{\xi}_i,\boldsymbol{\xi}_j).$
\cite{Johnson:Moore:Ylvisaker:1990}, \cite{Morris:etal:1993} and \cite{Mitchell:etal:1994} established that entropy optimal
designs for GPs are asymptotically equivalent to maximin designs as the
correlation becomes weaker, i.e. as $\rho\rightarrow 0$ in (\ref{eqn:corrfn}). % the
%correlation function $c(\boldsymbol{\xi},\boldsymbol{\xi}^\prime).$ 
This is a convenient result as typically the correlation parameter
$\rho$ is not known a-priori. % This asymptotic result provides a
%justifiable approach for designing experiments, particularly before data has been observed: one can use an initial space-filling design to approximate, or emulate, entropy optimal designs for GPs. Subsequently, the initial space-filling design can be sequentially updated to add additional design points using the entropy criterion with an updated estimate of the correlation parameter $\rho$ given the data collected so far.  
%
\begin{figure}[ht!]
	%	\figurebox{}{20pc}{}[Kfunction.pdf]
	\centering
	\captionsetup{width=.95\linewidth}
	\includegraphics[scale=0.5]{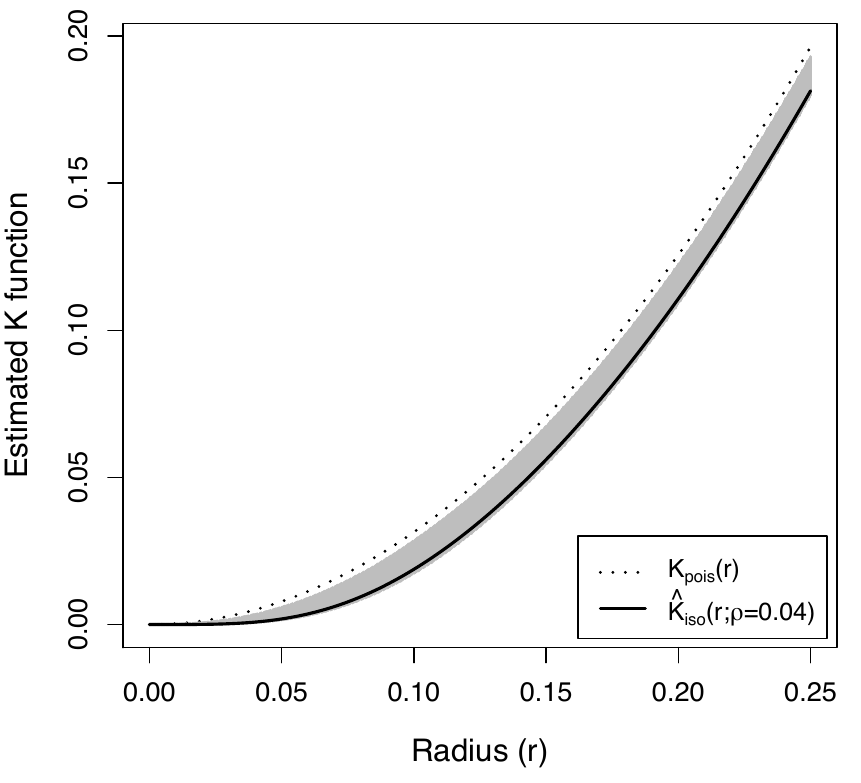}
	\caption{Estimated Ripley's $K$ function, $\hat{K}_{iso},$ taken as the pointwise median with corresponding 90\% uncertainty interval (shown in grey) from 1,000 replicates of estimated entropy-optimal designs (dotted line represents Ripley's function under random sampling, $K_{pois}$, for comparison).}
	\label{fig:Kfunc}
\end{figure}
To illustrate the connection between PPs %, space-filling designs, 
and entropy optimal designs for GPs, the following motivating simulation experiment is considered.
For reasons which will shortly become clear, entropy optimal designs for our simplified (stationary, isotropic) GP model will generate regular point patterns. 
When a PP is stationary and isotropic with constant first-order intensity $\lambda_Z$, one can use Ripley's $K$ function \citep{Ripley:1976,Diggle:etal:2010} to measure spatial dependence in the point patterns, defined to be
%\begin{eqnarray*}
$K_Z(r) = {E_Z(r)}/{\lambda_Z},$
%\end{eqnarray*}
where $E_Z(r)$ is the expected number of points within a distance of radius $r$
of an arbitrary point.
%
%\begin{figure}[t]
%\begin{center}
%\includegraphics{sim_dpp.pdf}
%\end{center}
%\caption{Estimated $K$ function with the corresponding 95\%
%  uncertainty interval from 100 replicates (dashed line represents the $K_R$ function under random sampling for comparison).}
%\label{sim:example}
%\end{figure}
%
The simulation experiment proceeds by generating 1,000 replicates of entropy optimal designs of size $n=30$ in $d=2$ dimensions.  Each of these entropy optimal designs is constructed by starting with a random initial design which is then optimized using the Fedorov Exchange Algorithm for 20,000 iterations.  The candidate set of input settings, $\chi,$ is taken to be the grid of $50\times 50$ equally spaced points in $[0,1]^2.$  The assumed exponential correlation function is $c(\boldsymbol{\xi}, \boldsymbol{\xi}^\prime) = 0.45^{ {||\boldsymbol{\xi}-\boldsymbol{\xi}^\prime||}_1}$ where $\vert\vert \cdot\vert\vert_1$ is the $L_1$ distance metric.
Figure \ref{fig:Kfunc} shows the estimated $K_Z$ function \citep{Diggle:etal:2010} averaged over the replicates and 90\% pointwise uncertainty intervals for the generated optimal designs as a function of radius $r$.  As a reference, the $K_R$ function for a completely random (Poisson process) point pattern is also shown as the dashed line in the figure.  
As the estimated $K_Z$ function is less than the $K_R$ function, this indicates evidence of a regular structure to the point pattern \citep{Ripley:1977,Dixon:2014}.

The evident connection between entropy optimal designs and a regular point pattern is relevant to our goal of emulating designs because of a connection between entropy optimal designs and a particular stochastic process model for a regular PP, which we explore in the next section.

% % % % % % % % % %
% % % % % % % % % %
% % % % % % % % % %
% % % % % % % % % %

%We propose a model-based perspective for optimal design.  Consider the process $Z$ as a generalization of the $N$-run measure $\xi_N$ defined in Section \ref{sec:introduction}.  The process $Z$ will define non-zero mass on all ${\bf x} \in \boldsymbol{\chi}$.  This definition has two interpretations.  The first is that $Z$ does not merely represent the (single) optimal design but rather all possible designs that could be constructed in $\boldsymbol{\chi}$.  Second, in the case of Bayesian design, we might think of $Z$ as also representing the possible designs under the prior distribution $\pi(\beta).$\\
%
%We would like to arrive at a process $Z$ from which samples of optimal designs could be drawn.  One could imagine explicitly arriving at the distribution of this process by theoretical means; that is given a model and criterion of interest, a measure on the space of designs $\Xi$ could potentially be expressed.  However, given the complexity of this problem, such an approach would not yield much benefit from the usual approach. Instead, we propose using design model based on point processes (PP) {\bf CITES}.  The idea is to find a suitably flexible point process model that shares important characteristics with optimal designs for a wide variety of statistical models.\\

% ======================================================================

\section{Optimal Design Emulator}
\label{sec:demu}
% Optimal Design Emulator

While the criterion-based view of optimal designs introduced in Section \ref{sec:introduction} is the popular interpretation in the literature, a more formal probabilistic  exposition %\citep{Kiefer:1985,Muller:2007} 
\citep{franco2008exploratory,franco:2008}
was previously explored.  
%The idea is as follows.  Define the set
%\[S_N=\left\lbrace
%\begin{tabular}{c}
%$p_1,p_2,\ldots,p_n$\\ % & ${\bf 0}$\\
%${\bf x}_1,{\bf x}_2,\ldots,{\bf x}_n$ % & $\chi\setminus\lbrace {\bf x}_1,\ldots,{\bf x}_n\rbrace$
%\end{tabular}\right\rbrace
%\]
%
%\noindent
%to be the exact design of the experiment where $N=\vert\chi\vert.$  The discrete design is then specified as a continuous function of ${\bf x}$, $Z({\bf x})$ such that $E[Z({\bf x}_i)]=p_i$ which satisfies the discrete probability measure
%$\sum_{{\bf x}\in\chi}E[Z({\bf x})]=1$,
%such that $E[Z({\bf x})]>0$ for  ${\bf x}\in S_N$ and $E[Z({\bf x})]=0$ for ${\bf x}\in\chi\backslash S_N$.  Here, $\chi\subset\mathbb{R}^d$ is a finite set of support points \citep{Kiefer:1985}, and the optimum results in the $p_i$'s placing all their weight on the points, ${\bf x}$'s, that minimize our criterion, with the interpretation being that ``the weights may be regarded as precision or duration of the measurements'' \citep{Muller:2007}.
Given such a probabilistic design model, one can imagine our design problem as simply being equivalent to sampling %, that is, drawing 
a realization of a point pattern distributed according to our model.  %In the earlier probabilistic exposition from the design literature, this probability model places non-zero weight only on the optimal design points.  
Formally, we can think of this as a conditional distribution, i.e. conditional on $\mathcal{L}(\boldsymbol{\xi}_1,\ldots,\boldsymbol{\xi}_n)=\mathcal{L}^*$ where
%where the conditionality arises from the criterion function of the design, $\mathcal{L}(\boldsymbol{\xi}_1,\ldots,\boldsymbol{\xi}_n)$ achieving a particular value, $\mathcal{L}^*,$ where
$\mathcal{L}^*=\min_{\boldsymbol{\xi}_1,\ldots.\boldsymbol{\xi}_n} \mathcal{L}(\boldsymbol{\xi}_1,\ldots,\boldsymbol{\xi}_n).$  

More broadly, we can cast the design problem as one of needing to define a probability model for any finite collection of points of cardinality $n$ which could make up our design.  %Unconditionally, we can imagine a corresponding probability model for a fixed cardinality, $n$, of the resulting point pattern.  
That is, unconditionally our model will define the probability of all $n$-run point patterns over the (discrete) sample space $\chi.$  The optimal design is one (or a small subset, say, due to isometries) of the point patterns which collectively form the sample space.  Denote the stochastic process generating these point patterns by $Z$, let $f_Z$ represent the probability mass (or density) function of this process, and let $\mathcal{J}(\boldsymbol{\xi}_1,\ldots,\boldsymbol{\xi}_n)=\mathcal{M}\left(\mathcal{L}(\boldsymbol{\xi}_1,\ldots,\boldsymbol{\xi}_n)\right)$ for some monotonically decreasing transformation $\mathcal{M}$ such that $\mathcal{J}$ is a non-negative, Lebesgue-integrable function of the inputs.
\vspace{12pt}
%{\bf Definition:}
\begin{definition}
\label{def:emulator}
We call the probability model represented by the mass  function $f_Z$ a {\em design emulator} if $f_Z \propto \mathcal{J},$ where $f_Z$ and $\mathcal{J}$ are defined on the same support $\chi.$
\end{definition}
\vspace{12pt}
%
%Note that this notion of a design emulator is in terms of the probability representation of a design.  In other words, we aim to introduce a statistical model to emulate the probability of the proposed designs. Leveraging this alternative representation, our task will be to arrive at an appropriate emulator for the probability of design points and to use this model as a means of sampling the optimal design from the stochastic PP $Z$ without resorting to the brute-force optimization techniques that are prevalent in optimal design.
Note that Definition \ref{def:emulator} relates the original loss criterion specification of optimal design to a probability mass function which arrives at the same solution by taking the arg max (mode) or maximum a posteriori (MAP) solution of the probability mass function with respect to the design points.

In order to construct a PP-based design emulator for our GP regression model of interest, we are motivated by the so-called {\em determinantal} PP model from the PP literature, which we introduce in Section \ref{sec:dpp}.  This model generates point patterns that fall in the regular pattern class of point processes, as introduced in Section \ref{sec:ppod}.  Later, we will explore a variant of this model which will motivate a computationally cheap algorithm for emulating the optimal design from our stochastic PP model.

\subsection{Determinantal Point Processes}
\label{sec:dpp}
An increasingly popular PP model that generates regular point patterns is the {\em determinantal point process (DPP)}.  The DPP was introduced to the statistics literature only recently \citep{Hough:etal:2006,Kulesza:etal:2011,lavancier:2014}.  It has been applied to sparse variable selection problems~\citep{Rock:George:2015,Mutsuki_2016} and statistical and machine learning~\cite[e.g.,][]{kulesza_2013_determinantal,kang2013fast,affandi2014learning,dupuy2016learning, Xu_2016}.

For entropy optimal designs of GP models, the following result on discrete, finite DPPs due to \cite{kulesza_2013_determinantal} motivates the use of DPP models.
%
%\vspace{.3cm}
%\noindent{\bf Lemma 1.}~\citep{kulesza_2013_determinantal}
\begin{lemma}\label{lemma3}~\citep{kulesza_2013_determinantal}
For a DPP defined over a discrete candidate set $\chi\subset\mathbb{R}^d$ with the positive semi-definite kernel function
$K(\boldsymbol{\xi},\boldsymbol{\xi}^\prime;\boldsymbol{\theta})$ with $\boldsymbol{\xi},\boldsymbol{\xi}^\prime\in\chi$ and known kernel function
parameter $\boldsymbol{\theta},$ the probability mass function for a point pattern realization $Z$ of cardinality $n$ is given by
$f_Z\propto \det({\bf K}_Z)$,
where the $n\times n$ positive semi-definite matrix ${\bf K}_Z$ has entries ${[{\bf K}_Z]}_{ij}=K(\boldsymbol{\xi}_i,\boldsymbol{\xi}_j;\boldsymbol{\theta}).$
\end{lemma}
Based on this result, we immediately have the following.
\begin{corollary}
%\noindent{\bf Corollary.} 
The entropy-optimal design for GP regression model (\ref{eqn:gpmodel}) with $\sigma^2=1,\sigma_\epsilon^2=0$ and correlation function $c(\cdot,\cdot;\boldsymbol{\rho})$ corresponds to the mode of a DPP with kernel function $K(\boldsymbol{\xi},\boldsymbol{\xi}^\prime;\boldsymbol{\theta})\equiv c(\boldsymbol{\xi},\boldsymbol{\xi}^\prime;\boldsymbol{\rho})$ conditional on having cardinality $n$.
\end{corollary}
%
%\vspace{.3cm}
%
These results provide an elegant connection between entropy optimal designs for GP regression and using DPP models to essentially emulate the point pattern associated with the optimal design by placing a DPP prior on the space of point patterns to which the optimal design belongs.  However, on the surface, finding the mode of the DPP is no easier than the usual optimization problem associated with finding the entropy optimal design.
A key result, due to \cite{Hough:etal:2006}, leads to the following approximation to the DPP, known as the {\em Determinantal Projection Point Process} (DPPP), defined over a set of $N$ candidate points $\chi$.
%
%\noindent
%{\bf Lemma 2.}~\citep{Hough:etal:2006} 
\begin{lemma}~\citep{Hough:etal:2006} 
Suppose $Z$ is a DPP with kernel $K$ defined over $\chi$ and write
$K(\boldsymbol{\xi},\boldsymbol{\xi}^\prime)=\sum_{k=1}^N\lambda_k\phi_k(\boldsymbol{\xi})\phi_k^T(\boldsymbol{\xi}^\prime)$
where $\phi_k$'s are orthonormal eigenvectors of $K$ with eigenvalues
$\lambda_k>0\ (k=1,\ldots,N),$ such that $\lambda_k>\lambda_{k^\prime}$ for $k<k^\prime.$  
Define $\mathcal{Z}$ to be a DPPP with kernel given by $\mathcal{K}(\boldsymbol{\xi},\boldsymbol{\xi}^\prime)=\sum_{k=1}^NB_k\phi_k(\boldsymbol{\xi})\phi_k^T(\boldsymbol{\xi}^\prime)$ where the $B_k$'s follow independent  $\textrm{Bernoulli}({\lambda_k}/(\lambda_k+1))$ distributions for $k=1,\ldots,N.$  Then $\mathcal{Z} {\buildrel d \over =} Z,$ where ${\buildrel d \over =}$ denotes equality in distribution.
\end{lemma}
\noindent
This result shows that any DPP can be represented as the weighted combination of  DPPP's.  \citet{Hough:etal:2006} show that this result implies a sampling algorithm where one first generates the Bernoulli random variables $B_1,\ldots,B_N$ where the number of points in the realization is $n=\sum_{i=1}^NB_i$, and then the locations of the points are generated by (suitably orthonormalized) vectors, whose $L_2$ norm is interpreted as a discrete probability measure.  In other words, sampling from the DPP is simplified by the separation of {\em how many points} make up a realization and the {\em location of points} for a realization of a given size; that is
\vspace{-0.25cm}
\begin{eqnarray*}
&& \hspace{-1cm} P\left(\boldsymbol{\Xi}=\lbrace \boldsymbol{\xi}_1,\ldots,\boldsymbol{\xi}_n\rbrace,B_1=b_1,\ldots,B_N=b_N \right)\\
&=& P\left(\boldsymbol{\Xi}=\lbrace \boldsymbol{\xi}_1,\ldots,\boldsymbol{\xi}_n\rbrace\middle|  B_1=b_1,\ldots,B_N=b_N\right)P\left(B_1=b_1,\ldots,B_N=b_N\right),
\end{eqnarray*}
where $\sum_{i=1}^NB_i=n$ is the total number of points appearing in a particular realization.

For our purposes, the generation of point patterns of random cardinality is not relevant, however the approximation introduced by the DPPP gives us the tools to specify a conditional framework that eventually can be used to give an (approximate) emulator of the entropy optimal design.

\subsection{Fixed Rank Determinantal Point Process}
\citet{Kulesza:etal:2011} outline the notion of a fixed-rank DPP, a DPP with a fixed sample size $n$. That is, the sampling of the $B_i$'s is conditional on $\sum_{i=1}^NB_i=n$.  While technically elegant, their approach is less interpretable from a statistical modeling perspective.  In our approach, we recognize the distribution of $B_1,\ldots,B_N$ given  $\sum_{i=1}^N B_i=n$ as a {\em conditional Bernoulli distribution} \citep{Chen:Liu:1997}.  The advantage of our approach is two-fold.  First, we can define a clear hierarchical statistical model for sampling from a fixed-rank DPP.  Second, \cite{Chen:Liu:1997} provide no less than four algorithms for sampling from this conditional Bernoulli distribution, with differing computational and memory complexity tradeoffs.  This allows one to provide a more efficient algorithm for constructing entropy optimal designs.

Conditioning on $\sum_{i=1}^N B_i=n$, we have the following hierarchical model for the fixed-rank DPP, 
\begin{eqnarray*}
&& \hspace{-1cm} 
P\left(\boldsymbol{\Xi}=\lbrace\boldsymbol{\xi}_1,\ldots,\boldsymbol{\xi}_n\rbrace, B_1=b_1,\ldots,B_N=b_N\middle|
  \sum B_i=n\right) \\
&=& P\left(\boldsymbol{\Xi}=\lbrace\boldsymbol{\xi}_1,\ldots,\boldsymbol{\xi}_n\rbrace\middle| \lbrace B_j=1\rbrace_{j\in\mathcal{S}},\lbrace B_j=0\rbrace_{j\in\lbrace 1,\ldots,N\rbrace\backslash\mathcal{S}}\right) \;\; \times \;\\
&&  P\left(\lbrace B_j=1\rbrace_{j\in\mathcal{S}},\lbrace B_j=0\rbrace_{j\in\lbrace 1,\ldots,N\rbrace\backslash\mathcal{S}}\ \middle|\ \vert\mathcal{S}\vert=n\right),
\end{eqnarray*}
\noindent
where $\mathcal{S}$ is the subset of indices of $\lbrace
1,\ldots,N\rbrace$ of cardinality $\vert\mathcal{S}\vert=n$.
Calculation of the first term comes from $L_2$-norms of appropriate
orthonormalizations of the vectors as shown in Algorithm 1.
The conditional Bernoulli probability can be calculated sequentially as
\begin{eqnarray*}
&& \hspace{-1cm}
 P\left(\lbrace B_j=1\rbrace_{j\in\mathcal{S}},\lbrace B_j=0\rbrace_{j\in\lbrace 1,\ldots,N\rbrace\backslash\mathcal{S}}\middle| \sum_{i=1}^N B_i=n\right)\\
&=& P\left(B_1=b_1\middle| \sum_{i=1}^N B_i=n\right)\times P\left(B_2=b_2\middle| B_1=b_1,\sum_{i=1}^N B_i=n\right)  \times \cdots \times \\
& & P\left(B_N=b_N\middle| B_1=b_1,\ldots,B_{N-1}=b_{N-1},\sum_{i=1}^N B_i=n\right),
\end{eqnarray*}
using the recursive  method of \cite{Chen:Liu:1997} which is summarized in the Supplementary Material.  Generating a fixed-rank DPP realization using conditional Bernoulli sampling then proceeds similarly as in \citep{Hough:etal:2006}, as shown in Algorithm~\ref{alg1}.  That is, we first select the $n$ eigenvectors according to the conditional Bernoulli sampling, as parameterized by the eigenvalues $\lambda_1,\ldots,\lambda_N.$  Then, given the $n$ selected eigenvectors we draw the design by iteratively drawing each design point to be added, $\boldsymbol{\xi}_j,$ according to the conditional discrete probability distribution $P_j(\boldsymbol{\xi}),$ which is updated each time a new design point is added to the design.  This continues until all $n$ design points have been drawn.  

Note that ${\bf K}_\chi$ denotes the kernel matrix constructed for the candidate set $\chi.$  Given the eigendecomposition of ${\bf K}_\chi,$ the computational complexity of Algorithm~\ref{alg1} is dominated by the calculation of $P(j,r)$ which is $\mathcal{O}(nN)$ using the ID-checking-sampling algorithm of \cite{Chen:Liu:1997}, giving the overall algorithm an order of $\mathcal{O}(nN^2)$ in the worst case (although unlikely since generation of $\mathcal{S}$ is probabilistic, not deterministic).  In practice, Algorithm 1 may itself be dominated by the eigendecomposition of ${\bf K}_\chi$ when $N$ is large.
%
%\begin{minipage}[c]{1.0\textwidth}
%\begin{spacing}{1.5}
%\begin{lstlisting}[frame=lrb,numbers=none,framerule=1pt,mathescape=true,label=alg:1,framexleftmargin=-4pt,framexrightmargin=2pt,caption=Generating a fixed-rank DPP realization.]
% Input: $\phi_1,\ldots,\phi_N$ and $\lambda_1,\ldots,\lambda_N$ from eigendecomposition of ${\bf K}_\chi=\sum\lambda_i\phi_i\phi_i^T$
%
% // Draw from the conditional Bernoulli distribution
% Set $\mathcal{S}_0=\lbrace\rbrace$ and $j=0$
% Repeat
%	Set $j=j+1$
%	Let $r=\vert \mathcal{S}_{j-1}\vert.$
%	With probability $P(j,r)$ (see Appendix)
%		Set $\mathcal{S}_j=\mathcal{S}_{j-1}\cup j$
% Until $\vert S_j\vert=n$
% Set $\mathcal{S}=\mathcal{S}_n$
%
% // Initialize required quantities given $\mathcal{S}$
% Let $v(\boldsymbol{\xi})=(\phi_{\mathcal{S}[1]}(\boldsymbol{\xi}),\ldots,\phi_{\mathcal{S}[n]}(\boldsymbol{\xi}))^T$
% Let $e_j$ be the vector of 0's except 1 in the $j$th position, $j=1,\ldots,n$
%
% // Draw the point pattern
% for $j$ in $n,\ldots,1$
%    Sample $\boldsymbol{\xi}_j$ from $P_j(\boldsymbol{\xi})=\frac{1}{j}\left(\vert\vert v(\boldsymbol{\xi})\vert\vert^2-\sum_{k=1}^{n-j}\vert e_k^Tv(\boldsymbol{\xi})\vert^2\right)$
%    Orthonormalize $\phi_1,\ldots,\phi_{j-1}$ with respect to $e_j$
%
% // Return the drawn fixed rank point pattern realization
% return $\boldsymbol{\Xi}_n=(\boldsymbol{\xi}_1,\ldots,\boldsymbol{\xi}_n)^T$
%\end{lstlisting}
%\end{spacing}
%\end{minipage}
%
\begin{algorithm}[!h]
\linespread{0}\selectfont
\caption{Generating a fixed-rank DPP realization.}\label{alg1}
\vspace{-0.2cm}
\enspace Input: $\phi_1,\ldots,\phi_N$ and $\lambda_1,\ldots,\lambda_N$ from eigendecomposition of ${\bf K}_\chi=\sum\lambda_i\phi_i\phi_i^T$\\
\enspace \texttt{// Draw from the conditional Bernoulli distribution}\\
\enspace Set $\mathcal{S}=\lbrace\rbrace$ and $j=0$\\
\enspace Repeat\\
\qquad \enspace	Let $r=\vert \mathcal{S}\vert.$\\
\qquad \enspace	Set $j=j+1$\\
\qquad \enspace	With probability $P(j,r)$ (see Appendix) \\
\qquad \qquad \enspace		Set $\mathcal{S}=\mathcal{S}\cup j$\\
\enspace Until $\vert S\vert=n$\\
%\enspace Set $\mathcal{S}=\mathcal{S}_n$\\
\enspace \texttt{// Initialize required quantities given $\mathcal{S}$}\\
\enspace Let $v(\boldsymbol{\xi})=(\phi_{\mathcal{S}_1}(\boldsymbol{\xi}),\ldots,\phi_{\mathcal{S}_n}(\boldsymbol{\xi}))^T$\\
\enspace Let $e_j$ be the vector of 0's except 1 in the $j$th position, $j=1,\ldots,n$\\
\enspace \texttt{// Draw the point pattern}\\
\enspace for $j$ in $n,\ldots,1$\\
\qquad \enspace    Sample $\boldsymbol{\xi}_j$ from $P_j(\boldsymbol{\xi})=\frac{1}{j}\left(\vert\vert v(\boldsymbol{\xi})\vert\vert^2-\sum_{k=1}^{n-j}\vert e_k^Tv(\boldsymbol{\xi})\vert^2\right)$\\
\qquad \enspace  Orthonormalize $\phi_1,\ldots,\phi_{j-1}$ with respect to $e_j$\\
\enspace \texttt{// Return the drawn fixed rank point pattern realization}\\
\enspace return $\boldsymbol{\Xi}_n=(\boldsymbol{\xi}_1,\ldots,\boldsymbol{\xi}_n)^T$\\
\end{algorithm}

\subsection{Emulating the Optimal Design}
Algorithm~\ref{alg1} will generate fixed-rank DPP realizations with a fixed number of points, and while these points should generally exhibit a regular pattern, there is no guarantee that any particular realization would be of especially high quality. %in terms of the regularity of the point pattern.  
That is, much like any stochastic process, it is always possible to draw a ``bad'' realization that has low probability.

Since the entropy optimal design for GP regression corresponds to the mode of a DPP by the corollary to Lemma~\ref{lemma3}, this suggests relating the optimal design problem to the DPP sampling problem.  
Note that the optimal design problem can be written in terms of the candidate matrix involved in DPP sampling, since $K(\boldsymbol{\xi},\boldsymbol{\xi}^\prime)=\sum_{k=1}^N\lambda_k\phi_k(\boldsymbol{\xi})\phi_k^T(\boldsymbol{\xi}^\prime)$ and $R(\boldsymbol{\xi},\boldsymbol{\xi}^\prime)=\sum_{i=1}^N\lambda_i\tilde{\phi}_i(\boldsymbol{\xi})\tilde{\phi}_i(\boldsymbol{\xi}^\prime)^T$ where $\tilde{\phi}_i=\left(\phi_{i,\sigma(1)},\ldots,\phi_{i,\sigma(n)}\right)$ extracts the desired $n$ entries of eigenvectors $\phi_i,\ i=1,\ldots,N,$ with the length-$n$ subset $\lbrace\sigma(1),\ldots,\sigma(n)\rbrace\subset \lbrace 1,\ldots,N\rbrace,$ such that $\sigma(j)\neq\sigma(k)\ \forall j\neq k.$
%$\tilde{\phi}=W^T\phi$ for some $N\times n$ matrix $W$ containing $0$'s and $1$'s such that each row has at most a single $1$ and each column sums to $1.$  Essentially, the matrix $W$ extracts the desired $n$ columns of $\phi$, and we can alternatively express these columns as $\sigma(1),\ldots,\sigma(n)\subset \lbrace 1,\ldots,N\rbrace.$  
The optimal design expressed in this way becomes $\left(\boldsymbol{\xi}_1,\ldots,\boldsymbol{\xi}_n\right) :=\left(\chi_{\sigma^*(1)},\ldots,\chi_{\sigma^*(n)}\right)$ where
\begin{eqnarray}
& &\left(\sigma^*_{(1)},\ldots,\sigma^*_{(n)}\right)=  \\
&&{\arg\max}_{\sigma(1),\ldots,\sigma(n)}\det\!\left(\sum_{i=1}^N\lambda_i\phi_i(\sigma(1),\ldots,\sigma(n))\phi_i^T(\sigma(1),\ldots,\sigma(n))\right),\nonumber
\end{eqnarray}
showing that the optimization in this form occurs explicitly in terms of the elements of the eigenvectors of the candidate matrix ${\bf K}_\chi.$
Since the eigenvalues are positive, this allows us to interpret each eigenvalue as being the weight given to each candidate point in the design by a given eigenvector, eigenvalue pair.
%Furthermore, since the eigenvectors are orthonormal, we know that each eigenvalue is less than or equal to one.  This allows us to interpret each eigenvalue as being the weight given to each candidate point in the design by a given eigenvector, eigenvalue pair.
%Furthermore, since the eigenvectors are orthonormal, we know that each eigenvector element is $\leq 1,$ lending them the interpretation of how much weight is given to each candidate point by a given eigenvector, eigenvalue pair.

It is tempting to simplify this problem by only considering the first $n$ eigenvector, eigenvalue pairs when performing the optimization.  As an empirical approach, this is  supported by the wide usage of low-rank kernel matrix approximations in GP regression, the simplest version of which extracts the first eigenpairs and discards the rest.  Additionally, there is a more direct connection motivated by the following result of \cite{ko:etal:1995}.  

\begin{lemma}~\citep{ko:etal:1995}
Let $\sum_{j=1}^n\alpha_j\zeta_j\zeta_j^T,$ where $\alpha_j>\alpha_{j^\prime}$ for $j<j^\prime,$ represent the eigendecomposition of the positive semi-definite matrix ${\bf R}$ for a proposed design $\boldsymbol{\xi}_1,\ldots,\boldsymbol{\xi}_n$ where the proposed design is a subset of the design candidates composing ${\bf K}_\chi$.  Let $\lbrace\lambda_j\rbrace$ be the eigenvalues of ${\bf K}_\chi,$ where $\lambda_j>\lambda_{j^\prime}$ for $j<j^\prime.$  Then, $\alpha_j\leq \lambda_j, j=1,\ldots,n.$
\end{lemma}

Lemma 3 immediately shows that $\prod_{j=1}^n\alpha_j\leq\prod_{j=1}^n\lambda_j,$ that is the determinant of a proposed $n-$ run design is bounded above by the product of the first $n$ eigenvalues of the candidate matrix with equality occuring when the eigenvector entries in $\phi$ place all of their weight on the proposed design points and zero elsewhere.  That is, for some canonical kernel having non-zero eigenvector entries placing weight on the optimal design points $\boldsymbol{\xi}_1,\ldots,\boldsymbol{\xi}_n$ in the first $n$ eigenvectors and the remaining $N-n$ candidate points $\chi\setminus \boldsymbol{\xi}_1,\ldots,\boldsymbol{\xi}_n$ having non-zero weight among the remaining candidate points, the solution to (7) using only the first $n$ eigenpairs nonetheless yields the optimal design. Motivated by these connections, in the context of our DPP design emulator, we propose to take the {\em Conditional Optimal Design} (COD),
%\vspace{-0.25cm}
\begin{eqnarray*}
\boldsymbol\Xi=\arg\max_{\boldsymbol{\xi}_1,\ldots,\boldsymbol{\xi}_n}P\left(\lbrace\boldsymbol{\xi}_1,\ldots,\boldsymbol{\xi}_n\rbrace \middle| \lbrace B_j=1\rbrace_{j\in\mathcal{S}^*},\lbrace B_j=0\rbrace_{j\in\lbrace 1,\ldots,N\rbrace\backslash\mathcal{S}^*}\right),
\end{eqnarray*}
where
%\vspace{-0.25cm}
\begin{eqnarray}
\label{eqn:sstar}
\mathcal{S}^* &=&\arg\max_{\mathcal{S}}P\left(\lbrace B_j=1\rbrace_{j\in\mathcal{S}},\lbrace B_j=0\rbrace_{j\in\lbrace 1,\ldots,N\rbrace\backslash\mathcal{S}}\middle| \sum_{i=1}^N B_i=n\right),
\end{eqnarray}
as our (approximate) emulator of the optimal design.  The proposed COD solution has theoretical support since it is the most probable conditional solution as shown by the following theorem.
\begin{theorem}\label{thrm3}
%{\bf Theorem 3.}  
The index set given by (\ref{eqn:sstar}) %$$\mathcal{S}^*=\arg\max_{\mathcal{S}}P\left(\lbrace B_j=1\rbrace_{j\in\mathcal{S}},\lbrace B_j=0\rbrace_{j\in\lbrace 1,\ldots,N\rbrace\backslash\mathcal{S}}\middle| \sum_{i=1}^N B_i=n\right)$$ 
is $\mathcal{S}^*=\lbrace 1,\ldots,n\rbrace.$ %$\frac{\lambda_j}{\lambda_j+1}, j=1,\ldots,n$.
\end{theorem}
%\noindent
%
%
%\begin{minipage}[c]{1.0\textwidth}
%\begin{spacing}{1.5}
%\begin{lstlisting}[frame=lrb,numbers=none,framerule=1pt,mathescape=true,label=alg:2,framexleftmargin=-4pt,framexrightmargin=2pt,caption=Drawing a fixed rank design from the design emulator.]
% Input: $\phi_1,\ldots,\phi_n$ from the eigendecomposition of ${\bf K}_\chi=\sum\lambda_i\phi_i\phi_i^T$
%
% // Initialize required quantities
% Let $v(\boldsymbol{\xi})=(\phi_{1}(\boldsymbol{\xi}),\ldots,\phi_{n}(\boldsymbol{\xi}))^T$
% Let $e_j$ be the vector of 0's except 1 in the $j$th position
%
% // Draw the point pattern
% for $j$ in $n,\ldots,1$
%    Set $\boldsymbol{\xi}_j = \arg\max_{\boldsymbol{\xi}} P_j(\boldsymbol{\xi})=\frac{1}{j}\left(\vert\vert v(\boldsymbol{\xi})\vert\vert^2-\sum_{k=1}^{n-j}\vert e_k^Tv(\boldsymbol{\xi})\vert^2\right)$
%    Orthonormalize $\phi_1,\ldots,\phi_{j-1}$ with respect to $e_j$
%
% // Return design drawn from the entropy optimal design emulator
% return $\boldsymbol{\Xi}_n=(\boldsymbol{\xi}_1,\ldots,\boldsymbol{\xi}_n)^T$
%\end{lstlisting}
%\end{spacing}
%\end{minipage}
%
Theorem~\ref{thrm3} shows that selecting the set $\mathcal{S}^*$ amounts to calculating the first $n$ (eigenvector, eigenvalue) pairs (i.e. such that $\lambda_1>\lambda_2>\ldots >\lambda_n$) of the $N\times N$ kernel matrix ${\bf K}_\chi$.  This eliminates the computational burden of conditional Bernoulli sampling, leading to a fast algorithm for emulating the optimal design.  In addition, finding  good low-rank approximate eigendecompositions of large matrices is a well-studied problem \cite[e.g.][]{halko:etal:2011}, removing the need to compute the full eigendecomposition in the practical application of Algorithm 2.  We refer to this fast sampling algorithm for the mode of the fixed-rank DPP as our {\em optimal design emulator} of entropy designs for GP regression.  
\begin{algorithm}[!h]
\linespread{0}\selectfont
\caption{Drawing a fixed rank design from the design emulator.}\label{alg2}
\vspace{-0.2cm}
\enspace Input: $\phi_1,\ldots,\phi_n$ from the eigendecomposition of ${\bf K}_\chi=\sum\lambda_i\phi_i\phi_i^T$\\
\enspace \texttt{// Initialize required quantities}\\
\enspace Let $v(\boldsymbol{\xi})=(\phi_{1}(\boldsymbol{\xi}),\ldots,\phi_{n}(\boldsymbol{\xi}))^T$\\
\enspace Let $e_j$ be the vector of 0's except 1 in the $j$th position\\
\enspace \texttt{// Draw the point pattern}\\
\enspace for $j$ in $n,\ldots,1$\\
\qquad \enspace  Set $\boldsymbol{\xi}_j = \arg\max_{\boldsymbol{\xi}} P_j(\boldsymbol{\xi})=\frac{1}{j}\left(\vert\vert v(\boldsymbol{\xi})\vert\vert^2-\sum_{k=1}^{n-j}\vert e_k^Tv(\boldsymbol{\xi})\vert^2\right)$\\
\qquad \enspace Orthonormalize $\phi_1,\ldots,\phi_{j-1}$ with respect to $e_j$\\
\enspace \texttt{// Return design drawn from the design emulator}\\
\enspace return $\boldsymbol{\Xi}_n=(\boldsymbol{\xi}_1,\ldots,\boldsymbol{\xi}_n)^T$\\
\end{algorithm}
%
%Note that in addition to the eigen-argument above, this also has the nice interpretation of the popular approach of simplifying kernel modeling methods by taking a low-rank approximation of the kernel matrix involving the first $n$ eigenvector-eigenvalue pairs.  This is known to give the best kernel matrix approximation in terms of Frobenius norm.  However, in the context of DPP sampling, Theorem\ref{thrm3} shows that this corresponds to the most probable sub-spectrum.

The proposed pseudo-code for drawing from the design emulator is shown in Algorithm~\ref{alg2}.  Note that the inputs to Algorithm~\ref{alg2} depend on the matrix ${\bf K}_\chi$ having been formed with a ``suitable'' value of the correlation parameter $\rho.$  As noted earlier, space-fillingness occurs as $\rho\rightarrow 0$ \citep{Mitchell:etal:1994}; in practice we choose a suitably small setting of $\rho$ to construct space-filling designs using the design emulator.

\begin{figure}
\centering
\captionsetup{width=.95\linewidth}
%\figurebox{10pc}{}{}[dppgaussianexample3.eps]
\includegraphics[scale=0.49]{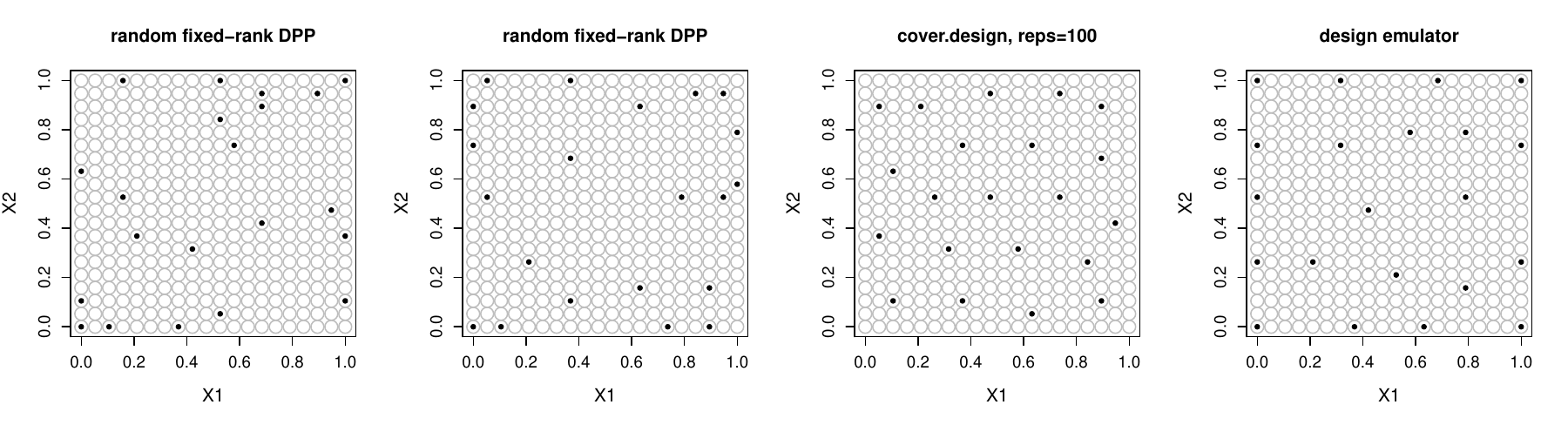}
\caption{Random realizations drawn from the fixed-rank DPP (left two panels), a space-filling design constructed by 100 random restarts of the \texttt{cover.design} function from \texttt{R} package \texttt{fields}, and the optimal design emulator constructed as the (approximate) mode of the fixed-rank DPP with isotropic Gaussian correlation function in $p=2$ dimensions with correlation parameter $\rho=0.01$.  All are $n=21$ point designs.}
\label{fig:dppdesign}
\end{figure}

An example of the design obtained from Algorithm~\ref{alg2} %the design emulator 
in 2 dimensions is shown in Figure \ref{fig:dppdesign} along with 2 random realizations of the fixed-rank DPP drawn according to Algorithm~\ref{alg1} and a design constructed using the
space-filling %design function 
\texttt{cover.design} of
\citet{Nychka:etal:2015}.  Comparisons in terms of the determinants, integrated mean-squared prediction error (IMSPE) and runtimes % of
%the correlation matrices for the resulting designs as well as their
%runtimes
are summarized in Table \ref{tab:desruns}.  %Since it is
%typically recommended to perform random restarts of
%\texttt{cover.design} to obtain a solution closer to the global
%optima, we report the runtime for 0 and 100 restarts.  
Note that
for any random realization of the fixed-rank DPP there is no guarantee that
the sampled points will be especially good in terms of
space-fillingness; here the two samples drawn seem poor.  However, the
design emulator results in a design that empirically fills the space
well and also produces the best determinant criterion value.  At the same time, the design emulator
was the fastest of all methods (even using unoptimized \texttt{R} code).  While running 100 restarts of \texttt{cover.design} improved the criterion, it also %significantly 
increases the computational cost.  As one might expect, the \texttt{cover.design} does provide the best IMSPE metric, but the design emulator is not too far behind, and is an order of magnitude better than the random DPP designs.  The difference comes down to how the design emulator and \texttt{cover.design} behave near the boundaries of the region, reflecting the differing goals of entropy-optimal designs versus IMSPE-optimal designs.
%
%\begin{figure}[ht!]
%\begin{center}
%\includegraphics[scale=0.49]{dppgaussian_example3.pdf}
%\end{center}
%\caption{Random realizations drawn from the fixed-rank DPP (left two panels), a space-filling design constructed by 100 random restarts of the \texttt{cover.design} function from \texttt{R} package \texttt{fields}, and the optimal design emulator constructed as the (approximate) mode of the fixed-rank DPP with isotropic Gaussian correlation function in $p=2$ dimensions with correlation parameter $\rho=0.01$.  All are $n=21$ point designs.}
%\label{fig:dppdesign}
%\end{figure}
%
\begin{table}[ht!]
\begin{center}
%\caption{Values of the determinant of the correlation matrix for designs resulting from 2 random samples of the fixed-rank DPP, 2 designs generated by \texttt{cover.design} using 0 and 100 random restarts, and the emulated optimal design taken as the (approximate) mode of the fixed-rank DPP.  The runtimes are indicated in seconds.}
\caption{Comparison of design algorithms}
\label{tab:desruns}
\vspace{-0.3cm}
\begin{tabular}{p{0.16\linewidth}|p{0.11\linewidth}p{0.11\linewidth}p{0.18\linewidth}p{0.18\linewidth}p{0.09\linewidth}}
\hline
Design\ \ \ \ \ \ \ \ \ \ \ \ Algorithm & random\ \ \ \ fixed-rank DPP & random\ \ \ \  fixed-rank DPP  & \texttt{cover.design} reps=0 & \texttt{cover.design} reps=100 & design\ \ \ emulator\\
\hline
\hline
%det$({\bf R}(\boldsymbol{\Xi}_n))$ & 2.15e-20 & 2.59e-19 & 1.71e-20 & 2.79e-20 & {\bf 5.24e-14}\\
$\log\det({\bf R}(\boldsymbol{\Xi}_n))$ & -45.28 & -42.79 & -45.51 & -45.02 & {\bf -30.57}\\
IMSPE          &  1.0e-2 & 1.3e-2 & 1.7e-3 & {\bf 1.5e-3} & 2.8e-3\\
Runtime (s)  & 312.0 & 441.5 & 0.5 & 37.8 & {\bf 0.2}\\
\hline
\end{tabular}
%\vspace{-0.5cm}
\end{center}
%\vspace{-0.2cm}
\end{table}

\subsection{Batch Sequential Design via Design Emulator}
The design drawn from the design emulator requires specification of the correlation function parameter, $\rho,$ of the GP.  Due to the emulator's speed, it becomes feasible to sample designs for different values of this parameter, or to sequentially update designs with refined estimates of $\rho$ as data are collected.  Here we will demonstrate a batch-sequential approach in the case of an isotropic GP.

From Lemma~\ref{lemma3}, we have $f_Z\propto \det({\bf K}_Z)$.  Suppose an initial design of size $n_1$ has already been selected consisting of locations $\boldsymbol{\Xi}_1=\left(\boldsymbol{\xi}_1,\ldots,\boldsymbol{\xi}_{n_1}\right)$ with the corresponding kernel sub-matrix ${\bf K}_{\boldsymbol{\Xi}_1}$ from the overall kernel matrix ${\bf K}_\chi$ defined on the candidate set $\chi.$  Then, 
\begin{align*}
f_Z  &= f_{Z\setminus \boldsymbol{\Xi}_1}\times f_{\boldsymbol{\Xi}_1}
%	&\propto det(K_{Z\setminus \boldsymbol{\Xi}_1})\times det(K_{\boldsymbol{\Xi}_1})\\
	\propto \det(\widetilde{{\bf K}}_{Z\setminus\boldsymbol{\Xi}_1})\times \det({\bf K}_{\boldsymbol{\Xi}_1})
\end{align*}
where $\det(\widetilde{{\bf K}}_{Z\setminus\boldsymbol{\Xi}_1})=\det({\bf K}_{Z\setminus\boldsymbol{\Xi}_1}-{\bf k}_{\boldsymbol{\Xi}_1,Z\setminus\boldsymbol{\Xi}_1}^T{\bf K}_{\boldsymbol{\Xi}_1}^{-1}{\bf k}_{\boldsymbol{\Xi}_1,Z\setminus\boldsymbol{\Xi}_1})$ denotes the Schur complement of ${\bf K}_Z$ in ${\bf K}_{\boldsymbol{\Xi}_1},$ with ${\bf K}_{Z\setminus\boldsymbol{\Xi}_1}$ being the submatrix of ${\bf K}_Z$ excluding rows, columns associated with the design points $\boldsymbol{\Xi}_1$ and ${\bf k}_{\boldsymbol{\Xi}_1,Z\setminus\boldsymbol{\Xi}_1}$ being the (rectangular) cross-kernel matrix between design points $\boldsymbol{\Xi}_1$ and the set of points $Z\setminus\boldsymbol{\Xi}_1.$ 

The above decomposition allows for a simple sequential updating scheme.  Suppose $n_1$ design points $\boldsymbol{\Xi}_1$ have been selected so far. % (perhaps in a one-shot arrangement, or perhaps as a result of some previous sequential design point selection iterations).  
To select $n_2$ additional design points, say $\boldsymbol{\Xi}_2$, one performs the following steps.
\begin{enumerate}
\item Construct $\widetilde{{\bf K}}_{\chi\setminus\boldsymbol{\Xi}_1}$.  This matrix assigns probability $0$ to (re)selecting any of the first $n_1$ points.
\item Apply Algorithm 2 using $\widetilde{{\bf K}}_{\chi\setminus\boldsymbol{\Xi}_1}$.  This conditionally samples the next $n_2$ points, $\boldsymbol{\Xi}_2.$ % conditional on already having selected the first $n_a$ points.
\item The updated design of $n=n_1+n_2$ points can then be returned as $\boldsymbol{\Xi}=\boldsymbol{\Xi}_1 \cup \boldsymbol{\Xi}_2$.
\end{enumerate}
%The sequential selection steps could then be iterated again to perform further updates.

This conditional approach to sequential design is very elegant.  We note two useful applications. %, and which will be demonstrated in Section 4.2.  
First, consider sequential designs for GP regression.  An initial design would be constructed using a small setting of $\rho$ to emulate a space-filling design.  Sequential design updates can then use a data-supported values of $\rho,$  %At each iteration, the kernel matrix can be updated using the most recent point estimate of $\rho$, thereby 
resulting in sequential designs that start off as space-filling but which evolve to extract more meaningful information from the process being observed.
%$ {\bf K}_\chi(\cdot,\cdot;\rho_0) \rightarrow \widetilde{{\bf K}}_{\chi\setminus\boldsymbol{\Xi}_1}(\cdot,\cdot;\rho_1)\rightarrow \widetilde{{\bf K}}_{\chi\setminus\boldsymbol{\Xi_1}\cup\boldsymbol{\Xi}_2}(\cdot,\cdot;\rho_2)\rightarrow \cdots,$
%where the initial ``space-filling'' value $\rho_0$ is subsequently updated based on the data collected. 

Another relevant scenario is the desire to enforce certain {\em projection properties} of designs.  Typically, it is desirable % the desired projection property is 
to enforce space-fillingness and/or non-collapsingness in all marginal dimensions as well as in the full $d$-dimensional design space. For instance, it is well known that LHS designs preserve space-fillingness of the 1-dimensional marginals but do not enforce this constraint on the higher-order marginals.  %Generally, adding this constraint to design construction has been a challenge, both in formulating an appropriate mathematical criterion and in optimizing the resulting criterion.  
Recently, \cite{joseph2015maximum} proposed a criterion that aims to preserve the space-fillingness constraint in all marginal sub-spaces when constructing $d$-dimensional designs, but in general there has been little work in this area due to the computational difficulty of finding such designs. 

Fortunately, for separable kernels (such as (\ref{eqn:corrfn})) the DPP was recently shown to preserve space-fillingness in all lower-dimensional subspaces in the continuous setting \citep{mazoyer:2019}.   
Space-fillingness of the lower-dimensional projections is a good property, but it could come at the expense of space-fillingness in higher dimensions if one focuses myopically only on the lower-dimensional properties.  On the other hand, focusing on higher-dimensional space-fillingness may cause lower dimensional projections to have duplicate coordinate values even though overall the points do exhibit space-fillingness in the lower-dimensional projection.  In other words, it is possible to produce a design that is more effective at simultaneously space-fiilling the lower and higher dimensional spaces.  The conditioning procedure we outline removes the possibility of overlapping points in lower-dimensional projections to increase the efficiency of lower-dimensional calculations.  So, we propose enforcing non-collapsingness via a conditioning constraint, resulting in designs with lower-dimensional projections that are both space-filling and non-overlapping.

Our sequential formulation allows one to easily enforce the non-collapsing projection property constraint -- that is, to remove the possibility of design points overlapping in their marginal projections.  %First, consider the $d$-dimensional point $\boldsymbol{\xi}=(\xi_1,\ldots,\xi_d)\in \boldsymbol{\Xi}$ that is already part of our design.  For any point not in the design, say $\boldsymbol{\xi}^\prime\notin\boldsymbol{\Xi}$, this point would violate the marginal projection property for dimension $j$ if $\xi^\prime_j=\xi_j$ for some $\boldsymbol{\xi}\in\boldsymbol{\Xi}.$  Similarly, $\boldsymbol{\xi}^\prime$ would violate the 2D marginal projection property in dimensions $j,k$ if $\xi^\prime_j=\xi_j\text{ and }\xi^\prime_k=\xi_k$ for some $\boldsymbol{\xi}\in\boldsymbol{\Xi}.$
Let \[S_1=\lbrace \boldsymbol{\xi}^\prime\in \chi\setminus \boldsymbol{\Xi} \ \vert\   \xi_j^\prime=\xi_j \text{ where } \boldsymbol{\xi}\in\boldsymbol{\Xi} \text{ for at least one } j\in 1,\ldots,d\rbrace,\]
\[S_2=\lbrace \boldsymbol{\xi}^\prime\in \chi\setminus \boldsymbol{\Xi} \ \vert\  \xi_j^\prime=\xi_j\cap \xi^\prime_k=\xi_k \text{ where } \boldsymbol{\xi}\in\boldsymbol{\Xi} \text{ for at least one pair } (j,k)\in 1,\ldots,d\rbrace,\]
\begin{align*}
S_3&=\lbrace \boldsymbol{\xi}^\prime\in \chi\setminus \boldsymbol{\Xi} \ \vert\  \xi_j^\prime=\xi_j\cap \xi_k^\prime=\xi_k\cap \xi_l^\prime=\xi_l\\ 
&\ \ \ \ \ \text{ where } \boldsymbol{\xi}\in\boldsymbol{\Xi} \text{ for at least one tuple } (j,k,l)\in 1,\ldots,d\rbrace,
\end{align*}
and so on, where the higher-order sets $S_4,\ldots,S_{d-1}$ are similarly defined.  Then, it is clearly the case that $S_1\supseteq S_2\supseteq\cdots\supseteq S_{d-1}.$
Therefore, %to find the subset of candidate points that would violate the desired  projection constraint in all marginal dimensions, 
it is sufficient to find the set $S_1$ alone.  This is an operation that is $\mathcal{O}(Nnd)$ in the worst case. % (where $N$ is the cardinality of $\chi$, $n$ is the number of design points and $d$ is the dimension of the input space). %, exhibiting no combinatorial explosion with dimension.  
Our batch-sequential design algorithm satisfying the non-collapsing constraint is outlined in Algorithm~\ref{alg3}.
%
%
%\begin{minipage}[c]{1.0\textwidth}
%\begin{spacing}{1.5}
%\begin{lstlisting}[frame=lrb,numbers=none,framerule=1pt,mathescape=true,label=alg:3,framexleftmargin=-4pt,framexrightmargin=2pt,caption=Batch-Sequential Design Emulator with Non-Overlapping Projections.]
% Let $\boldsymbol{\Xi}_a$ be the $n_a$ existing $d$-dimensional design points $\boldsymbol{\xi}_1,\ldots,\boldsymbol{\xi}_{n_a}$
% Let $\chi$ be the original (full) set of candidate points
% Initialize $S=\boldsymbol{\Xi}_a$
%
% // Construct the set of points violating the non-collapsing 
% // projection constraint of the existing design
% for $i$ in $1,\ldots,n_a$
%    for $j$ in $1,\ldots,d$
%        for $k$ in $1,\ldots,N$
%            if $\xi_{ij}==\chi_{kj}$ and $\chi_{k}\notin S$
%                $S=S\cup \chi_k$
%
% // Calculate the conditional kernel matrix
% $\widetilde{{\bf K}}_{\chi\setminus S}={\bf K}_{\chi\setminus S}-{\bf k}^T_{S,\chi\setminus S}{\bf K}^{-1}_{S}{\bf k}_{S,\chi\setminus S}$
%
% // Draw $n_b$ batch-sequential points using Algorithm 2 with the
% // first $n_b$ eigenfunctions $\phi_1,\ldots,\phi_{n_b}$ from the eigendecomposition
% // of kernel matrix $\widetilde{{\bf K}}_{\chi\setminus S}=\sum\lambda_i\phi_i\phi_i^T.$ 
%\end{lstlisting}
%\end{spacing}
%\end{minipage}
%
\begin{algorithm}[!h]
\linespread{0}\selectfont
\caption{Batch-Sequential Design Emulator with Non-Overlapping Projections.}\label{alg3}
\vspace{-0.2cm}
\enspace Let $\boldsymbol{\Xi}_a$ be the $n_a$ existing $d$-dimensional design points $\boldsymbol{\xi}_1,\ldots,\boldsymbol{\xi}_{n_a}$\\
\enspace Let $\chi$ be the original (full) set of candidate points\\
\enspace Initialize $S=\boldsymbol{\Xi}_a$\\
\enspace \texttt{// Construct the set of points violating the non-collapsing}\\
\enspace \texttt{// projection constraint of the existing design}\\
\enspace for $i$ in $1,\ldots,n_a$\\
\qquad \enspace for $j$ in $1,\ldots,d$\\
\qquad \qquad \enspace for $k$ in $1,\ldots,N$\\
\qquad \qquad \qquad \enspace if $\xi_{ij}==\chi_{kj}$ and $\chi_{k}\notin S$\\
\qquad \qquad \qquad \qquad \enspace $S=S\cup \chi_k$\\
\enspace \texttt{// Calculate the conditional kernel matrix}\\
\enspace $\widetilde{{\bf K}}_{\chi\setminus S}={\bf K}_{\chi\setminus S}-{\bf k}^T_{S,\chi\setminus S}{\bf K}^{-1}_{S}{\bf k}_{S,\chi\setminus S}$\\
\enspace \texttt{// Draw $n_b$ batch-sequential points using Algorithm 2 with}\\
\enspace \texttt{// the first $n_b$ eigenfunctions $\phi_1,\ldots,\phi_{n_b}$ from the}\\
\enspace \texttt{// eigendecomposition of kernel matrix $\widetilde{{\bf K}}_{\chi\setminus S}=\sum\lambda_i\phi_i\phi_i^T.$}\\
\end{algorithm}
%
% % % % % % % % % % % % % % % % % % % % % % % % % % % % % % %
% Scaling and sparsification approaches in next paper... :)
%
%\subsection{Sparse Approach for Scaling Design Emulation}
%The Wendland stuff goes here.  Maybe Nystrom also?

% ======================================================================

\section{Examples}
\label{sec:examples}
% Example Applications of the Design Emulator

%To motivate the interesting possibilities of using a fast design emulator, w
%We consider designs for Stochastic Gradient Descent (SGD) and GP regression. % space-filling fractional-factorial experiments, and {\bf A THIRD EXAMPLE?}.

\subsection{Designed Stochastic Gradient Descent}
%As mentioned earlier, one motivation for space-filling designs is as a variance-reduction technique when calculating statistical estimators.
%To demonstrate the potential for a computationally cheap design
%emulator for constructing space-filling designs in a modern context, we motivate possible modern
%applications in the big data setting involving the stochastic gradient
%descent (SGD) algorithm \citep{Bottou:2010}.  
SGD \citep{Bottou:2010} is used extensively in statistical machine learning to scale model training to big data for a variety of applications including linear models, clustering, GP regression and deep neural networks \citep[e.g.][]{Zhang:2004,Sculley:2010,dean2012large,Hensman:etal:2013,wan2013regularization,sutskever2014sequence,badrinarayanan2015segnet}.
SGD uses small random subsets of data, called {\em batches}, to estimate the gradient.  The idea of SGD is to sacrifice an increase in estimator variance for computational gain so the parameter space of the model can be more efficiently explored when fitting models to big data.  Due to the speed of the proposed
design emulator, we can replace SGD's random susbset selection with a
space-filling subset selection at each iteration of the algorithm, thereby recovering some of this variance tradeoff.

A small simulation was carried out by generating
observations from a $d=5$-dimensional linear regression model (similar to
the popular Friedman function \citep{Friedman:1991}),
\begin{align}
\label{eqn:friedman}
Y({\bf x})=\beta_0+\beta_1\sin(2\pi x_1x_2)+\beta_2(x_3-0.5)^2+\beta_3(x_4-0.5)^2+\beta_4x_4+\beta_5x_5+\epsilon,
\end{align}
where the regression coefficients $\beta_0,\ldots,\beta_5$ were 
generated as $\text{Unif}(-10,10)$ and the observational error was taken to be $\epsilon \sim N(0,1)$.  The model was fit using
SGD with batches of size $\texttt{batchsize}=\lbrace 23,43,63,83\rbrace$ and the model fit criterion was the usual sum of squared errors.  SGD iterates over all the batches of data in random order and repeats this entire process a number of times, called epochs.  We used $200$ epochs.
The total dataset size was $50\times \texttt{batchsize}$ and each study was replicated $100$ times, using randomly drawn coefficients for each replicate.  To evaluate the quality of the SGD solution, we compared the ratio of the average squared error of the regression coefficients for random subsets ($\text{MSE}(\hat{\beta}_j\vert\boldsymbol{\Xi}_R)$) versus  designed subsets ($\text{MSE}(\hat{\beta}_j\vert\boldsymbol{\Xi})$).  %For example, a ratio of 2 indicates the estimation error was twice as large as that of using SGD with the design emulator. 

The results summarized in Table \ref{tab:sgdexample} show that even in
such a simple example, the resulting error is usually 1.5-3 times
worse based solely on how the subsets are selected from the dataset.
This demonstrates a novel application of experimental design in the modern big data  setting that is enabled by the computationally cheap design emulator.
\begin{table}[t]
\begin{center}
\caption{Ratio of MSE of parameter estimates versus SGD subset size for the 5-dimensional test function given in Equation (\ref{eqn:friedman}). Random designs are denoted $\boldsymbol{\Xi}_R$ while optimal designs constructed using the emulator are denoted $\boldsymbol{\Xi}.$}
\label{tab:sgdexample}
\begin{tabular}{c|ccccc}
\hline
Subset Size & $\frac{\text{MSE}(\hat\beta_1\vert \boldsymbol{\Xi}_R)}{\text{MSE}(\hat\beta_1\vert \boldsymbol{\Xi})}$ & $\frac{\text{MSE}(\hat\beta_2\vert \boldsymbol{\Xi}_R)}{\text{MSE}(\hat\beta_2\vert \boldsymbol{\Xi})}$ & $\frac{\text{MSE}(\hat\beta_3\vert \boldsymbol{\Xi}_R)}{\text{MSE}(\hat\beta_3\vert \boldsymbol{\Xi})}$ & $\frac{\text{MSE}(\hat\beta_4\vert \boldsymbol{\Xi}_R)}{\text{MSE}(\hat\beta_4\vert \boldsymbol{\Xi})}$ & $\frac{\text{MSE}(\hat\beta_5\vert \boldsymbol{\Xi}_R)}{\text{MSE}(\hat\beta_5\vert \boldsymbol{\Xi})}$\\
\hline\hline
23     & 1.97 & 1.47 & 1.27 & 2.92 & 1.97\\
43     & 1.62 & 1.04 & 1.45 & 2.72 & 1.67\\
63     & 1.56 & 0.94 & 2.00 & 2.59 & 1.56\\
83     & 1.43 & 1.20 & 1.32 & 2.38 & 1.44\\
\hline
\end{tabular}
%\vspace{-0.3cm}
\end{center}
%\caption{Ratio of MSE of parameter estimates for random subsets ($\boldsymbol{\Xi}_R$) versus designed subsets ($\boldsymbol{\Xi}$) when using SGD to fit the model (\ref{eqn:friedman}).  Values greater than $1.0$ indicate the multiplicative factor by which random subsets had larger MSE relative to the designed subsets.}
%\label{tab:sgdexample}
%\vspace{-0.2cm}
\end{table}

\subsection{Batch-Sequential Designs in GP Regression}

\begin{figure}
\centering
\captionsetup{width=.95\linewidth}
%\figurebox{}{35pc}{}[3atatimeremoveprojupdaterho.eps]
%\figurebox{}{35pc}{}[4atatimeremoveprojupdaterho.eps]
\includegraphics[scale=0.47]{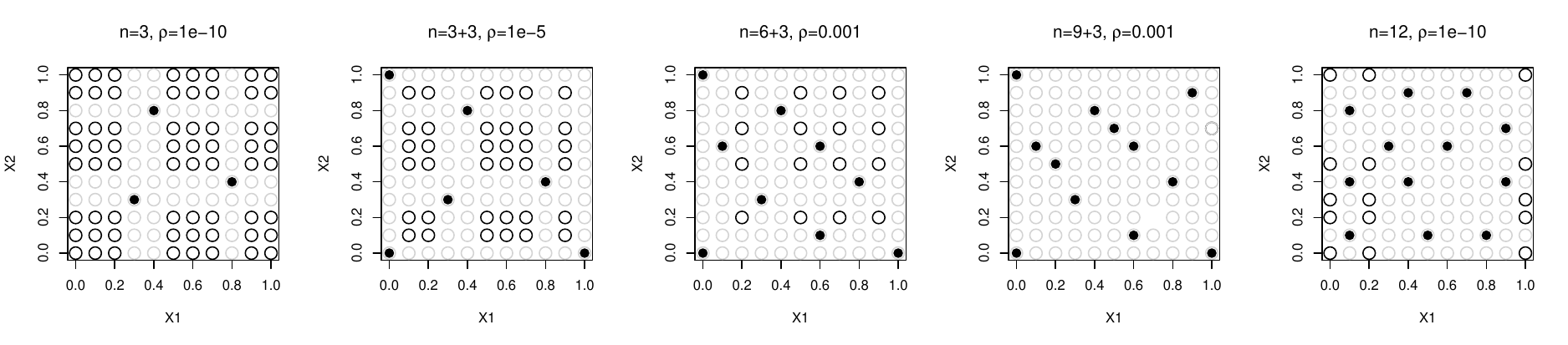}\\
\includegraphics[scale=0.47]{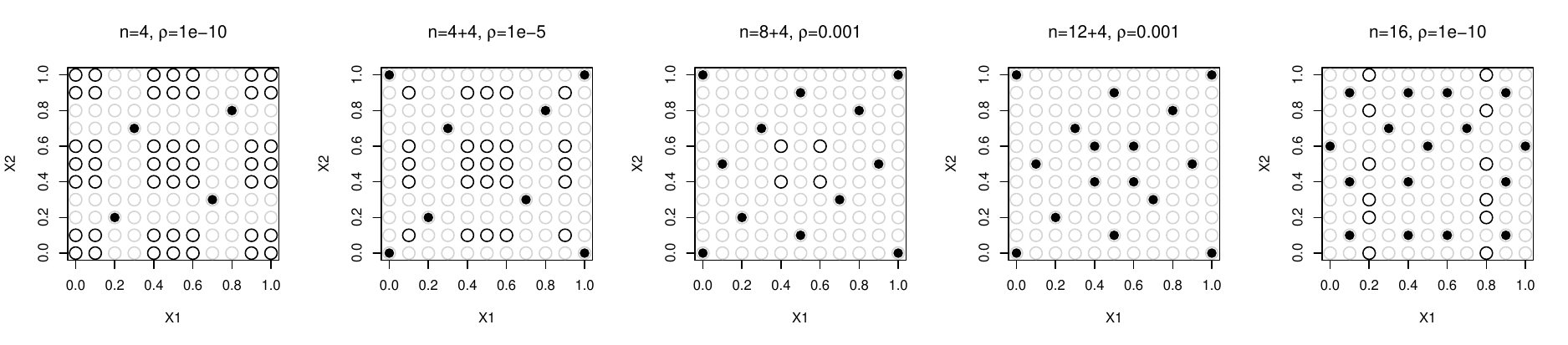}
\caption{Batch sequential designs with evolving $\rho$.  Row 1: batch size of 3. Row 2: batch size of 4. The single-shot designs ($n=12$ and $n=16$) are shown in the rightmost column.  Solid dots denote designs while empty circles are candidate points.  The gray circles denote candidates which would negatively impact the marginal projections.  These sequential designs were constructed to preserve the marginal projection constraint, and assume the correlation parameter is sequentially updated as $\rho=1\times 10^{-10} \rightarrow \rho=1\times 10^{-5} \rightarrow \rho=0.001 \rightarrow \rho=0.001$.  The single-shot designs ($n=12$ and $n=16$) were constructed using $\rho=1\times 10^{-10}.$}
\label{fig:3atatime3}
\end{figure}

Our second demonstration of the proposed technique considers constructing sequential designs for GP regression using Algorithm 3.  As outlined in Section 1.1, we assume a stationary GP regression model with mean zero and Gaussian correlation function with $\sigma_\epsilon^2=0.$  We take $d=2$ and consider batch-sequential designs constructed 3-at-a-time and 4-at-a-time as shown in Figure \ref{fig:3atatime3}.  In this figure, the design points are denoted by solid dots and the light gray circles denote available candidate points which, if selected, would negatively impact the space-fillingness of marginal projections. The black circles denote available candidate points that would not negatively impact the marginal projections.

In practice a sequential design benefits from improved estimates of the GP correlation parameter at each iteration. % from data collected at each step in the sequence.  
%Using Algorithm 3, we can construct designs which sequentially take this into account, becoming less space-filling while still preserving the desired non-collapsing marginal projection property.  
The sequential designs shown in Figure \ref{fig:3atatime3} reflect an evolution of $\rho$ over the 4 updates as $\rho=1\times 10^{-10}\rightarrow \rho=1\times 10^{-5} \rightarrow \rho=0.001 \rightarrow \rho=0.001.$  The sequences demonstrate a more centralized pattern to the points as the correlation is updated to represent an observed process that is smooth and slowly varying.  Yet, the marginal projection property is still satisfied, with no unoccupied marginal projections for the $n=9+3$ and $n=12+4$ designs as compared to the single-shot designs with $n=12$ and $n=16$ runs.  Designs that hold $\rho$ fixed and do not preserve the projection property are in the Supplementary Material.
%
%\begin{figure}[ht!]
%\centering
%\includegraphics[scale=0.47]{3atatime_removeproj_updaterho.pdf}\\
%\includegraphics[scale=0.47]{4atatime_removeproj_updaterho.pdf}
%\caption{Batch sequential designs with evolving $\rho$.  Row 1: batch size of 3. Row 2: batch size of 4. The single-shot designs ($n=12$ and $n=16$) are shown in the rightmost column.  Solid dots denote designs while empty circles are candidate points.  The gray circles denote candidates which would negatively impact the marginal projections.  These sequential designs were constructed to preserve the marginal projection constraint, and assume the correlation parameter is sequentially updated as $\rho=1\times 10^{-10} \rightarrow \rho=1\times 10^{-5} \rightarrow \rho=0.001 \rightarrow \rho=0.001$.  The single-shot designs ($n=12$ and $n=16$) were constructed using $\rho=1\times 10^{-10}.$}
%\label{fig:3atatime3}
%\vspace{0.5cm}
%\end{figure}

% ======================================================================

\section{Discussion}
\label{sec:discuss}
% Discussion

In this article we have introduced a novel probabilistic approach to constructing optimal designs motivated by PPs. We considered entropy-optimal designs, which have a clear connection to the popular space-filling designs used in computer experiments and spatial statistics, and we establish their connection to DPPs. By using a discrete version of this representation, we arrive at a computationally efficient algorithm to sample the mode of this PP, which corresponds to the optimal design.  Note that our methodology approximately samples the mode of this PP, yet in practice the quality of designs sampled resulted in criterion values many orders of magnitude larger than random DPP samples or space-filling LHS designs, which implies the approxmation is of high quality.  Subsequently, we extend the method to allow for sequential design construction and allow one to incorporate a popular marginal projection property. % without losing the computational benefits of our basic algorithm.  
%Since our approach to enforcing such constraints amounts to specifying a conditional kernel matrix, other constraints not considered in this paper could be easily implemented using the same basic approach described.

The design emulator introduced in this paper was demonstrated on the popular Stochastic Gradient Descent (SGD) algorithm by applying the emulator to fitting a variant of the popular Friedman test function where the data was generated using random coefficients. %SGD works by approximating the model gradient with a small batch sample from the entire dataset and is hugely popular in fitting complex statistical and machine learning models.  The algorithm trades off reduced computation time for increased variance in the estimation of the gradients.  
%We demonstrated that 
Using the emulator to design the SGD batches %from the design space 
noticeably improved the performance of SGD.
We also demonstrated the sequential variant of our algorithm for GP regression designs 
with non-collapsing marginal projections and evolving parameter estimates. % in a sequential model-fitting excercise.  
The designs constructed clearly show the effect of incorporating the marginal projection property constraint and the effect of updating the parameter as more data is collected. % This allows one to sequentially update the design, starting from a purely space-filling construction when we have not yet observed any information, to one which places greater focus on estimating the model as more data is collected.

Taking inspiration from earlier probabilistic designs in the literature, %\citep[e.g.][]{Kiefer:1985,Muller:2007}, 
this article provides a general approach to constructing designs from a PP perspective.  While we have focused on designs for stationary GP models, our approach would also apply to non-stationary GP models specified by a closed-form correlation function.  The methods outlined in this paper are available on \texttt{CRAN} in the \texttt{R}  package \texttt{demu}. %, and we aim to further develop this approach to handle more complex scenarios such as high-dimensional design construction and high-dimensional statistical learning algorithms.

% ======================================================================
\section*{Acknowledgement}
The work of Matthew T. Pratola was supported in part by the U.S. National Science Foundation (NSF) under Agreement NSF-DMS-1916231 and in part by the King Abdullah University of Science and Technology (KAUST) Office of Sponsored Research (OSR) under Award No. OSR-2018-CRG7-3800.3.  Peter F. Craigmile was supported in part by the NSF under grants NSF-DMS-1407604 and NSF-SES-1424481, and the National Cancer Institute of the National Institutes of Health under Award Number R21CA212308. The content is solely the responsibility of the authors and does not necessarily represent the official views of the National Institutes of Health.  C. Devon Lin's research was supported by the Discovery grant from the National Sciences and Engineering Research Council (NSERC) of Canada.

% ======================================================================
%\section*{Supplementary Material}
%
%Supplementary material available at \Bka\ online includes additional examples of classifying designs, calculation details for Algorithm 1, and further plots mentioned in 4.2.

\bibliographystyle{biometrika}
\bibliography{references}

%\appendix
\section*{Appendix}
\label{sec:app}

\section*{Proof of Corollary 1}
Given ${\bf \theta}$ (equivalently $\rho$) and taking $\mathcal{M}(\mathcal{L})=-\mathcal{L},$ the mode $\boldsymbol{\Xi}_n=\arg\max_{\boldsymbol{\xi}_1,\ldots,\boldsymbol{\xi}_n} f_Z$ necessarily minimizes (\ref{eqn:entopt}), and so the DPP mode is the optimal design.

\section*{Calculating $P(j,r)$ in Algorithm 1}
\citet{Chen:Liu:1997} derive the formula for $P(j,r)$ as well as the conditional Bernoulli sampler based on $P(j,r).$  
Let $C$ be any subset of $S_N=\lbrace 1,\ldots,N\rbrace$ and let $1\leq j \leq \vert C\vert.$
Let $S_j=\lbrace 1,\ldots,j\rbrace$ and $S_j^c=\lbrace j+1,\ldots,N\rbrace$.
For any $i\geq 1$, define $$T(i,C):=\sum_{j\in C}\lambda_j^i$$
and $$R(j,C):=\frac{1}{j}\sum_{i=1}^j(-1)^{i+1}T(i,C)R(j-i,C),$$
where $R(0,C):=1$ and $R(j,C)=0$ for any $j>\vert C\vert.$
Then,
$$P(j,r):=\frac{\lambda_j R(n-r-1,S_j^c)}{R(n-r,S_{j-1}^c)}.$$

\section*{Proof of Theorem 1}
Let ${\bf K}_\chi$ represent the positive semi-definite matrix defined over the set of candidate design points $\chi$, with corresponding eigenvalues $\lambda_1>\lambda_2>\cdots>\lambda_N.$  Let ${\bf B}=(B_1,\ldots,B_N)$ and 
let $\mathcal{B}_n=\lbrace {\bf b}=(b_1,\ldots,b_N): b_i\in\lbrace 0,1\rbrace \text{ and } \sum_{i=1}^Nb_i=n\rbrace$. \cite{Chen:Liu:1997} show that $P({\bf B}={\bf b})$ is given by
%\begin{align*}
$
P({\bf B}={\bf b}) = \prod_{k=1}^NP(k,\sigma_k)^{b_k}(1-P(k,\sigma_k)^{1-b_k}
\propto \prod_{k=1}^Nw_k^{b_k},
$
%\end{align*}
where $\sigma_k=\sum_{j=1}^kb_j$ with $\sigma_0=0$, $P(k,\sigma_k)$ is as defined above and $w_k=p_k/(1-p_k)$ with $p_k$ being the (independent, unconditional) probability of success for the $k$th Bernoulli.  But from \cite{Kulesza:etal:2011}, this is just $p_k=\frac{\lambda_k}{\lambda_k+1}$, so substituting we have $P({\bf B}={\bf b})\propto \prod_{k\in\mathcal{S}}\lambda_k.$  But, since the $\lambda_k$'s are monotone decreasing, $\prod_{k\in\mathcal{S}}\lambda_k>\prod_{k\in\mathcal{S}\setminus j}\lambda_k\lambda_l$ for any $l>j$ which implies $\prod_{k=1}^n\lambda_k>\prod_{k\in\mathcal{S}}\lambda_k$ for any $\mathcal{S}\neq \lbrace 1,\ldots,n\rbrace.$

%{\bf Proof:}  This amounts to showing that $\frac{\lambda_j}{1+\lambda_j}/\frac{\lambda_{j+1}}{1+\lambda_{j+1}}>1.$  But since the $\lambda_j$'s are monotone decreasing $\forall j$, take $\lambda_{j+1}=c_j\lambda_j$ for some $0<c_j<1.$  Substituting, we have $(1+c_j\lambda_j)/(c_j+c_j\lambda_j)>1,$ as required.

\end{document}